\documentclass[pra,aps,superscriptaddress,twocolumn,showpacs,nofootinbib]{revtex4-1}

\usepackage{placeins}
\usepackage{amsfonts, amsmath,amssymb,amsthm,bm}
\usepackage{mathrsfs}
\usepackage{graphicx}
\usepackage[usenames,dvipsnames]{color}
\usepackage{palatino}
\usepackage{fancyhdr}
\usepackage{mathrsfs}
\usepackage{multirow}
\usepackage{hyperref}
\usepackage{bigints}

\hypersetup{
   colorlinks=true,
   linkcolor=blue,
   citecolor=blue,
   urlcolor=blue
}

\newcommand{\tr}{\mathrm{tr}}
\newcommand{\R}{\mathbb{R}}

\newcommand{\N}{\mathbb{N}}
\newcommand{\id}{\mathbf{1}}

\theoremstyle{plain}
\newtheorem{theorem}{Theorem}

\newtheorem{lemma}[theorem]{Lemma}

\theoremstyle{definition}

\begin{document}
\title{Quantum computation is the unique reversible circuit model for which bits are balls}
\author{Marius Krumm}
\email{marius.krumm@univie.ac.at}\thanks{corresponding author.}
\affiliation{Institute for Quantum Optics and Quantum Information, Austrian Academy of Sciences, Boltzmanngasse 3, A-1090 Vienna, Austria}
\affiliation{Vienna Center for Quantum Science and Technology (VCQ), Faculty of Physics, University of Vienna, Boltzmanngasse 5, A-1090 Vienna, Austria}
\author{Markus P. M\"uller}
\email{markusm23@univie.ac.at}
\affiliation{Institute for Quantum Optics and Quantum Information, Austrian Academy of Sciences, Boltzmanngasse 3, A-1090 Vienna, Austria}
\affiliation{Perimeter Institute for Theoretical Physics, Waterloo, ON N2L 2Y5, Canada}

\begin{abstract}
The computational efficiency of quantum mechanics can be defined in terms of the qubit circuit model, which is characterized by a few simple properties: each computational gate is a reversible transformation in a connected matrix group; single wires carry quantum bits, i.e.\ states of a three-dimensional Bloch ball; states on two or more wires are uniquely determined by local measurement statistics and their correlations. In this paper, we ask whether other types of computation are possible if we relax one of those characteristics (and keep all others), namely, if we allow wires to be described by $d$-dimensional Bloch balls, where $d$ is different from three. Theories of this kind have previously been proposed as possible generalizations of quantum physics, and it has been conjectured that some of them allow for interesting multipartite reversible transformations that cannot be realized within quantum theory. However, here we show that all such potential beyond-quantum models of computation are trivial: if $d$ is not three, then the set of reversible transformations consists entirely of single-bit gates, and not even classical computation is possible. In this sense, qubit quantum computation is an island in theoryspace.
\end{abstract}

\date{January 28, 2019}

\maketitle

\section{Introduction}
Since the discovery of quantum algorithms that outperform all known classical ones in certain tasks~\cite{Shor}, improving our understanding of the possibilities and limitations of quantum computation has become one of the central goals of quantum information theory. While it is notoriously difficult to prove unconditional separation of polynomial-time classical and quantum computation~\cite{AaronsonBook}, an approach that is often regarded more tractable is to analyze how certain modifications of quantum computing affect its computational power. For instance, one may consider restrictions on the set of allowed quantum resources, and ask under which condition the possibility of universal quantum computation is preserved despite the restriction. Notable results along these lines, among many others, include the Gottesman-Knill theorem~\cite{GottesmanKnill,AaronsonGottesman,NielsenChuang}, insights on the necessity of contextuality as a resource for magic state distillation~\cite{Howard}, or bounds on the noise threshold of quantum computers~\cite{GottesmanIntro}.

In a complementary and in some sense more radical approach, going back to Abrams and Lloyd~\cite{Abrams}, one considers modifications of the quantum formalism itself and studies the impact of those modifications on the computational efficiency, resembling strategies of classical computer science such as the introduction of oracles~\cite{Sipser}. For example, it has been shown that availability of closed timelike curves leads to implausible computational power~\cite{AaronsonWatrous}, that stronger-than-quantum nonlocality reduces the set of available transformations~\cite{Barrett2007,Gross,AlSafi,Richens}, that tomographic locality forces computations to be contained in a class called AWPP~\cite{LeeBarrett,Barrett2017}, and that in some theories (satisfying additional axioms) higher-order interference  does not lead to a speed-up in Grover's algorithm~\cite{LeeSelby}. Further examples can be found e.g. in \cite{BoundsProofsAdviceGPTs,GarnerInteferometry,KickBack,OraclesGPT}.

\begin{figure}[!hbt]
\includegraphics[angle=0, width=.35\textwidth]{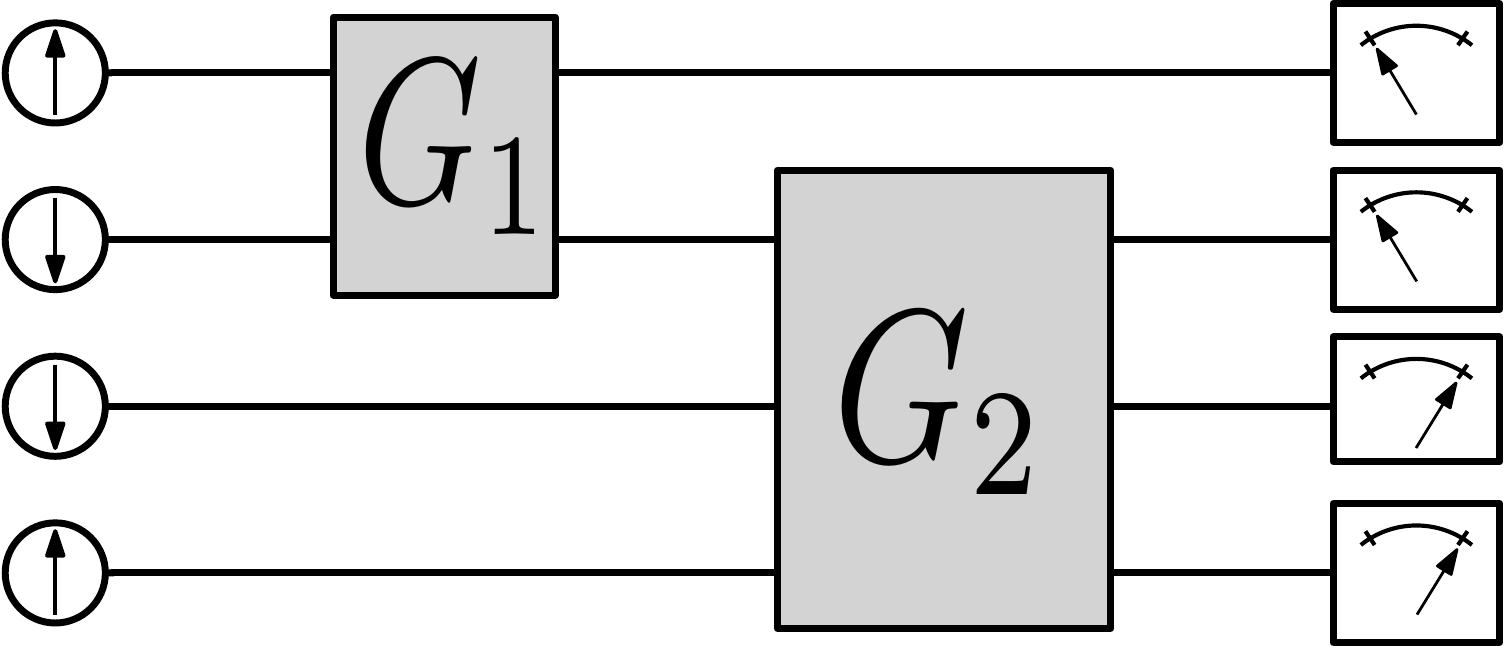}
\caption{The circuit model that we consider in this paper. We have an arbitrary finite number $n$ of wires (here $n=4$), and each wire carries a ``gbit'' which is a state in a $d$-dimensional Bloch ball state space. Initially, a product state is prepared (encoding, for example, the classical input to the algorithm), then a finite number of gates $G_i$ is applied, each acting on an arbitrary number of gbits, and finally local measurements are performed. We assume that the $G_i$ are elements of an (arbitrary unspecified) closed connected matrix group, and that the global state of $n$ wires is uniquely determined by the statistics and correlations of single-wire measurements (``tomographic locality''). If $d=3$, i.e.\ if the gbits are qubits, it has been shown in~\cite{Torre} that these assumptions uniquely characterize unitary quantum computation as the only computationally non-trivial theory. Here we analyze the case $d\neq 3$, and prove that --- despite conjectures to the opposite~\cite{DakicBrukner} --- the corresponding models do not allow for any non-trivial computation at all. We do \emph{not} assume that wires can be swapped, or that all transformations can be composed out of two-gbit transformations. See the main text for details.}
\label{fig_circuit}
\end{figure}

In this paper, we consider a specific modification of the quantum formalism that is arguably among the simplest and most conservative possibilities. This modification dates back to ideas by Jordan, von Neumann, and Wigner~\cite{JvNW}, and it has several independent motivations as we will explain further below. This generalization keeps all characteristic properties of quantum computation unchanged, but modifies a single aspect: namely, it allows the quantum bit to have any number of $d\geq 2$ degrees of freedom, instead of standard quantum theory's $d=3$ (or the classical bit's $d=1$). It has been conjectured~\cite{DakicBrukner} that the resulting theories allow for interesting ``beyond quantum'' reversible multipartite dynamics, which would make the corresponding models of computation highly relevant objects of study within the research program mentioned above. However, here we show that, quite on the contrary, these models are so constrained that they do not even allow for classical computation; hence, in Aaronson's terminology, the $d=3$ case of the standard qubit circuit model can be seen as an ``island in theoryspace''~\cite{Aaronson}.

Our paper is organized as follows. Section~\ref{SecGeneralized} gives the mathematical framework. We define single bits that generalize the qubit (``gbits''), and then give three postulates that allow us to reason about circuits that are constructed out of $n$ of these gbits. We formulate the problem that is addressed in this work and describe how it relates to earlier results in the literature. In Section~\ref{SecMainResult}, we state and prove our main result: namely, while our principles uniquely determine quantum computation in the case that the single gbits have $d=3$ degree of freedom, any other value of $d$ does not even allow for classical computation. We give the full proof for the case $d\geq 4$ (the $d=2$ case is deferred to the appendix), and illustrate the main idea of some of the proof steps by a circuit diagram, before concluding in Section~\ref{SecConclusions}.

\section{Generalized bits and gbit circuits}
\label{SecGeneralized}
In both classical and quantum computation, we can restrict our attention to the circuit model (as in Figure~\ref{fig_circuit}) where each of the wires (the single systems that enter and exit logical gates) corresponds to a two-level system. Quantum two-level systems (qubits) are different from classical ones (bits): they allow for a more complex behavior which encompasses phenomena like coherent superposition, interference, or uncertainty relations. Yet, both classical and quantum bits can be formalized in a unified way that we now describe (for both single and multiple bits, i.e.\ circuits, we follow the constructions and notation from~\cite{Torre}).

\subsection{Single gbits}
\label{SubsecSingleGbits}
To any $d\in\N$, we associate a ''generalized bit'' (gbit) that has the $d$-dimensional Bloch ball, $B^d=\{\vec a \in \R^d\,\,|\,\, |\vec a|\leq 1\}$, as its state space. Every vector $\vec a$ in the Bloch ball $B^d$ corresponds to a possible state of the generalized bit. Two-outcome measurements are described by vectors $\vec b\in\R^d$ with $|\vec b|=1$, such that the probability of the first outcome if performed on state $\vec a$ is $(1+\vec a \cdot \vec b)/2$, and that of the second outcome is $(1-\vec a \cdot \vec b)/2$. In the following, it will be convenient to use the notation $v(\vec a)=(1,\vec a)^\top\in\R^{d+1}$, such that these two probabilities become $\frac 1 2 v(\vec a)\cdot v(\pm \vec b)$. Reversible transformations of states are given by $\vec a \mapsto R\vec a$, where $R\in{\rm SO}(d)$ is a rotation matrix. These transformations map states to states and can be inverted (by applying $R^{-1}$), hence we can interpret them as closed-system time evolutions or, equivalently, reversible gates on single generalized bits.

For $d=3$, this formalism recovers the qubit of standard quantum theory~\cite{NielsenChuang}: as is well-known, every $2\times 2$ density matrix $\rho$ can be written in the form
\[
   \rho=(\mathbf{1}+\vec a_\rho \cdot \vec\sigma)/2,
\]
where $\vec\sigma=(\sigma_x,\sigma_y,\sigma_z)$ denotes the Pauli matrices. It is automatic in this representation that $\tr\,\rho=1$, and positivity $\rho\geq 0$ is equivalent to $|\vec a_\rho|\leq 1$. Hence the set of states of a quantum bit can be represented by the Bloch ball $B^3$. This representation has the important property that \emph{statistical mixtures correspond to convex combinations}: if a state $\rho$ is prepared with probability $p$ and another state $\rho'$ is prepared with probability $1-p$, then the total state $p\rho+(1-p)\rho'$ corresponds to the Bloch vector $\vec a_{p\rho+(1-p)\rho'}=p \vec a_\rho+(1-p)\vec a_{\rho'}$. This statistical interpretation of convex mixtures is also taken for balls of other dimensions $d\neq 3$, hence these Bloch balls can be regarded as state spaces of generalized probabilistic theories~\cite{Barrett2007}.

In the $d=3$ case, projective measurements are represented by unit vectors $\vec b$, $|\vec b|=1$, with outcome probabilities $(1\pm \vec a \cdot \vec b)/2$ as described above. Unitary transformations $U$ on states, acting as $\rho\mapsto U\rho U^\dagger$, are described in the Bloch ball picture by orthogonal maps $R_U$, $R_U^\top R_U=\mathbf{1}$, such that $\vec a_{U\rho U^\dagger} = R_U \vec a_\rho$. More general measurements (positive operator-valued measures) or transformations (completely positive maps) can also be described in the Bloch ball representation, but they are not needed in what follows and therefore omitted.

The simplest case of $d=1$ corresponds to the classical bit: there are two possible configurations, $\vec a=+1$ and $\vec a'=-1$, and further states that represent classical uncertainty about the configuration. Namely, if we have $+1$ with probability $p$ (and thus $-1$ with probability $1-p$), this corresponds to the state $p\vec a +(1-p)\vec a'$ in the interior the one-dimensional ``Bloch ball''.

There is one peculiarity in the $d=1$ case: instead of ${\rm SO}(1)=\{\mathbf{1}\}$, we should allow the group ${\rm O}(1)=\{-\mathbf{1},\mathbf{1}\}$ as Bloch ball transformations such that also the bit flip is allowed.

What is the significance of the $d$-dimensional Bloch balls if $d$ is neither one nor three? These gbits have appeared in various places in quantum information theory and the foundations of quantum mechanics. Historically, they have first shown up as precisely those two-level state spaces that can be described as (formally real, irreducible) Jordan algebras~\cite{JvNW}, a natural algebraic generalization of standard quantum theory. In fact, quantum theory with real amplitudes, i.e.\ over the field $\mathbb{R}$ instead of $\mathbb{C}$, has a $(d=2$)-dimensional Bloch ball as its ``quantum bit'', and the bits of quaternionic and octonionic quantum theory correspond to $B^d$ for $d=5$ and $d=9$ respectively. Furthermore, the fact that a two-level system should have a Euclidean ball state space can be derived from a variety of different sets of natural assumptions. In many reconstructions of quantum theory from physical or information-theoretic principles~\cite{Hardy,DakicBruknerQM,MasanesMueller,Chiribella,Hardy2011,HoehnToolbox,HoehnWever,Hoehn, Wilce, Coecke, Wetering}, this fact is derived as a first step. For example, postulating that the group of reversible transformations acts transitively on the pure states implies that the pure states must all lie on the unit hypersphere of an invariant inner product. If some points on the sphere were \emph{not} valid states, then there would exist additional measurements that would violate further natural postulates like Hardy's~\cite{Hardy} ``Subspaces'' axiom. This argumentation or others along similar lines~\cite{Hardy,DakicBruknerQM,MasanesMueller,Chiribella,Hardy2011,HoehnToolbox,HoehnWever,Hoehn,Goyal,Wilce, Coecke, Wetering} lead to Euclidean balls as the most natural state spaces of a generalized bit.

A more geometrical motivation can be found by considering spin-$\frac{1}{2}$ particles (compare e.g.\ to~\cite{DakicBrukner}): under rotations $\mathrm{SO}(3)$, they transform via $\mathrm{SU}(2)$. The density matrix transforms under the adjoint representation, which means that the Bloch vectors transform via the same rotation as in physical space. Therefore, the Bloch vector $\vec b$ can be seen as defining an oriented axis in physical space. The model considered in this paper is a direct generalization of the Bloch ball and this interpretation to arbitrary spatial dimensions. Indeed, the possibility that space might have more than three dimensions has appeared in a large variety of physical theories, see e.g. \cite{MoreDim1,MoreDim2,MoreDim3,MoreDim4,MoreDim5,MoreDim6}. It has also been argued that these generalized bits can be interpreted as ``information quasiparticles'' in some sense~\cite{Pawlowski}. In summary, these gbits are among the simplest and most natural generalizations of the classical bit and the qubit of quantum mechanics.

\subsection{Several gbits and computation}
\label{SubsecSeveralGbits}

To describe circuit computation, we need to define the state space, measurements, and transformations of several gbits. In standard quantum theory, where the gbits are qubits, there is a unique definition of these notions: the states of $n$ qubits are exactly the $(2^n)\times (2^n)$ density matrices, the reversible transformations are the unitaries, and the measurements are described by collections of projection operators. Similar definitions apply to $n$ classical bits. But if the gbits are Bloch balls of dimension $d\not\in\{1,3\}$, then it is apriori unclear what the composite state space should be.

Since we would like to be as general as possible, we will not make any attempt to fix the composite state space from the outset. Instead, we will work with a small set of principles that the composite $n$-gbit system is supposed to satisfy. While these principles will constrain the $n$-gbit state space, it is by no means obvious that they determine it uniquely. However, we will show below that they are indeed constraining enough to allow us to derive the full set of states and transformations.

An important principle is the \textbf{no-signalling principle}~\cite{Barrett2007}: \emph{the outcome statistics of measurements on any group of gbits does not depend on any other operations (e.g.\ measurements) that are performed on the remaining gbits.} This is a physically well-motivated constraint that lies at the heart of what we mean by ``different wires'' (i.e.\ subsystems) of the circuit in the first place.

This principle is satisfied by classical as well as quantum computation, and so is our second postulate of \textbf{tomographic locality}~\cite{Araki,Hardy}: \emph{every state on $n$ gbits is uniquely characterized by the statistics and correlations of the local gbit measurements.} In other words, a global $n$-gbit state is nothing but a catalog of probabilities for the outcomes of all the single-gbit measurements and their correlations.

It is not only classical and quantum theory that satisfies the principle of tomographic locality, but also more general probabilistic theories like boxworld~\cite{Gross}. If this principle was violated, then a collection of gbits would in some counterintuitive sense be ``more'' than a composition of its building blocks. Even though this formulation makes tomographic locality sound very natural, there are simple examples of theories that violate it. One such example is given by quantum theory over the real numbers $\mathbb{R}$~\cite{BGWshort,BGW}. This is because observables of two single real qubits do not linearly generate all observables of two real qubits. In particular, if $\sigma_y$ is the Pauli matrix with purely imaginary entries, then $\sigma_y$ is not a real qubit observable, but $\sigma_y\otimes\sigma_y$ is a real two-qubit observable. Intuitively, it represents a novel ``holistic'' degree of freedom that cannot be constructed out of local degrees of freedom and their correlations.

Not only is the postulate of tomographic locality very intuitive, but it is also very powerful: it allows us to represent states of $n$ gbits as \emph{tensors}~\cite{Barrett2007}. That is, even if we do not know what the set of $n$-gbit states is, we know that every such state can be written as an element of the linear space $(\mathbb{R}^{d+1})^{\otimes n}$ (in the quantum case, where $d=3$, this amounts to the $4^n$-dimensional real linear space of Hermitian $(2^n)\times(2^n)$ matrices; for real bits, it is the $2^n$-dimensional space that contains the probability vectors over $2^n$ configurations). In particular, an $n$-gbit product state with local Bloch vectors $\vec a_1,\ldots\vec a_n$ is represented by
\[
   v(\vec a_1,\ldots,\vec a_n):=(1,\vec a_1)^\top\otimes\ldots\otimes (1,\vec a_n)^\top,
\]
and all other states $\omega$ are vectors on the same space (but not of this product form). Tomographic locality then amounts to the fact that all these states are uniquely determined by the numbers
\[
   2^{-n} v(\vec b_1,\ldots, \vec b_n)^\top \omega,
\]
which are the outcome probabilities of local gbit measurements corresponding to the Bloch vectors $\vec b_1,\ldots,\vec b_n$ on the state $\omega$. This mathematical property has many intuitively appealing consequences that are not otherwise guaranteed, e.g.\ the property that products of pure states are pure. It is also the reason why the mathematical literature has focused almost entirely on this notion of composite state space (cf.\ e.g.~\cite{Namioka}): it leads to notions of ``tensor products'' of ordered linear spaces that allow one to prove general statements that are otherwise unavailable. In the context of this paper, it would seem extremely difficult to make any meaningful statements whatsoever if not even the linear space on which the global states live could be fixed from the outset. 

We need one further ingredient to arrive at a model of computation, namely a set of reversible transformations. In analogy to standard quantum computation (where these are the unitaries), we postulate that \textbf{the transformations form a closed connected matrix group}, and thus Lie group, $\mathcal{G}$: they form a group since they can be composed; they must be linear maps since if we prepare a state $\omega$ with probability $p$ and $\omega'$ with probability $(1-p)$, they must act on the components of the convex combination $p\omega+(1-p)\omega'$ individually, to be consistent with the probabilistic interpretation~\cite{Barrett2007}. Moreover, it is physically meaningful to model the group as closed since whenever we can approximate a transformation to arbitrary accuracy by gates, it makes sense to declare this transformation as in principle implementable.

This postulate is almost, but not quite, satisfied by classical computation, i.e.\ the $d=1$ case. As Bennett has shown~\cite{Bennett}, classical computation can be made fully reversible, at only marginal cost of space or time resources. There are finite universal gate sets (including e.g.\ Toffoli gates) that generate the full group of permutations of the $2^n$ configurations of the $n$ bits. These permutations therefore constitute the reversible transformations of the classical bits, and they form a closed matrix group of linear maps. This group, however, is discrete and not connected.

This discreteness is already reflected in the fact that the one-dimensional ``Bloch ball'' is discrete, i.e.\ has only a finite number (two) of pure states. Since the set of classical configurations (pure states) of $n$ bits is discrete, the group of reversible transformations must also be discrete. In the case $d\geq 2$ to which we thus restrict our attention in the following, however, even single bits (Bloch balls) contain a continuous manifold of pure states. In order to allow every pure state to evolve into every other (which we would expect to be crucial for the exploitation of the full computational potential), it is therefore necessary that the reversible transformations form a continuous group $\mathcal{G}$ --- in more detail, that $\mathcal{G}$ is a matrix Lie group such that its connected component at the identity is non-trivial. It then makes sense to consider continuous time evolution that implements elements of this connected component (as it is the case in quantum theory), and to disregard the mathematical possibility of having additional disconnected components. This motivates the assumption that $\mathcal{G}$ is connected.

All gates in a circuit will be elements of $\mathcal{G}$. This group must in particular contain the local qubit rotations: for $R\in{\rm SO}(d)$, write $\hat R(1,\vec a)^\top:=(1,R\vec a)^\top$, then the subgroup of local transformations is
\[
   \mathcal{G}_{\rm loc}:=\{\hat R_1\otimes \hat R_2\otimes\ldots\otimes \hat R_n\,\,|\,\, R_i\in{\rm SO}(d)\}.
\]
Note that we have used tomographic locality in deriving this prescription: since a local transformation acts like a product of transformations on the product states, it must act like this on all other states too since they live on the vector space that is spanned by the product states. Tomographic locality hence enforces that we can represent any linear map $X:(\R^{(d+1)})^{\otimes n}\to (\R^{(d+1)})^{\otimes n}$ as a tensor with $n$ upper and $n$ lower indices; that is,
\[
   X^{\alpha_1 \alpha_2\ldots \alpha_n}_{\beta_1 \beta_2\ldots \beta_n
   }:=(\vec e_{\beta_1}\otimes\ldots\otimes \vec e_{\beta_n})^\top X (\vec e_{\alpha_1}\otimes \ldots \otimes \vec e_{\alpha_n}),
\]
where $0\leq \alpha_i,\beta_i \leq d$, and $\vec e_\gamma$ denotes the $\gamma$-th unit vector, e.g.\ $\vec e_0=(1,0,\ldots,0)^\top$. This is in contrast to Bloch vectors $\vec b\in\R^d$, where we use the notation $\R^d\ni \vec b =\vec e_1=(1,0,\ldots,0)^\top$.

We demand that $\mathcal{G}_{\rm loc}\subseteq \mathcal{G}$, but do not make any further assumptions on $\mathcal{G}$. In particular, we do not assume that the $n$ gbits play physically identical roles: our assumptions allow in principle composite state spaces of $n$ gbits that are not symmetric with respect to permutations of the gbits. Hence we are also \emph{not} assuming that gbits can be reversibly swapped, or that other natural choices of transformations such as extensions of classical reversible gates (like CNOT) can necessarily be implemented. Therefore, our framework does not rely on the same set of assumptions as the circuit framework of \emph{symmetric monoidal categories}~\cite{CoeckeKissinger} that is often used in the quantum foundations context.

\subsection{The trivial case $\mathcal{G}=\mathcal{G}_{\rm loc}$}
For any Bloch ball dimension $d$, there is a trivial computational model: namely the choice that $\mathcal{G}=\mathcal{G}_{\rm loc}$. This describes a theory where the \emph{only} possible reversible transformations are independent local transformations of the single gbits. Such a model does not even allow for classical gates like the CNOT; it only admits gates and computations that evolve the gbits independently from each other without ever correlating them, i.e.\ products of single-gbit gates. A state space that is compatible with this choice of global transformations is simply
\[
   {\rm conv}\left\{(1,\vec a_1)^\top \otimes \ldots \otimes (1,\vec a_n)^\top\,\,|\,\, \vec a_i\in B^d \right\},
\]
i.e.\ all convex combinations of product states. This is a state space that does not contain entanglement.

\subsection{$d=3$ equals quantum computation, and relation to earlier work}
For the case of the standard qubit, i.e.\ of $d=3$, it has been proven in~\cite{Torre} that there is only a single possible non-trivial ($\mathcal{G}_{\rm loc}\subsetneq \mathcal{G}$) theory that satisfies the assumptions from above: namely, standard quantum theory over $n$ qubits, with the $(2^n)\times(2^n)$ density matrices as the states, and the projective unitary group $\mathcal{G}={\rm PU}(2^n)$ of transformations. That is, the postulates on composition of gbits from above, together with the structure of the single qubit, are sufficient to determine qubit quantum computation uniquely.

While this result is interesting in its own right, it is also the main motivation for the present work: if quantum computation is characterized by such a simple list of principles, then maybe one obtains other interesting models of computation by slightly tweaking one of the postulates. Since large parts of the mathematical structure are determined by the postulates on composition (no-signalling and tomographic locality), the most promising road towards modifying the setup and also keeping important mathematical tools seems to be to modify the structure of the single qubit --- and technically as well as conceptually (as explained in Subsection~\ref{SubsecSingleGbits}), the most natural way to do this is by changing the dimension of the Bloch ball $d$.

In the special case of $n=2$ gbits, the consequences of the above postulates have been explored in~\cite{MMAPG1,MMAPG2}. There it has been proven that \emph{the only consistent choice of transformations for Bloch ball dimension $d\neq 3$ is given by the trivial choice $\mathcal{G}=\mathcal{G}_{\rm loc}$.} However, computation is typically taking place on a large number $n\gg 2$ of gbits, and the techniques of~\cite{MMAPG1,MMAPG2} cannot readily be generalized to $n>2$.

In fact, it has been suggested in~\cite{DakicBrukner} that it is essential for Bloch ball dimensions $d\geq 4$ to allow for genuine $m$-partite interaction of the gbits, where $m\geq d-1\geq 3$.  Without a conclusive proof or explicit construction of the state space, the authors conjectured that interesting multipartite reversible dynamics is possible for such systems. In contrast to quantum theory, this $m$-partite dynamics would not be decomposable into two-gbit interactions. While tomographic locality has not been assumed in~\cite{DakicBrukner}, it is an important first step to check their conjecture under this additional assumption. In fact, it has been argued in~\cite{MM3D} that in the context of spacetime physics (the Bloch balls are interpreted in~\cite{DakicBrukner} as carrying some sort of $d$-dimensional spin degrees of freedom), tomographic locality is to be expected due to arguments from group representation theory.

This gives us another, independent motivation to ask the main question of this paper: \emph{if $d\neq 3$ and $n$ is any finite number of gbits, then what are the possible theories that satisfy the assumptions of Subsection~\ref{SubsecSeveralGbits}?}

\section{Main result}
\label{SecMainResult}
The main result of this work is an answer to the question posed at the end of the previous section:
\begin{theorem}
\label{TheMain}
Consider a theory of $n$ gbits, where single gbits are described by a $(d\geq 2)$-dimensional Bloch ball state space, subject to the single-gbit transformation group ${\rm SO}(d)$. As described above, let us assume no-signalling, tomographic locality, and that the global transformations form a closed connected matrix group $\mathcal{G}$.

If $d\neq 3$, then necessarily $\mathcal{G}=\mathcal{G}_{\rm loc}$, i.e.\ the only possible gates are (independent combinations of) single-gbit gates. No transformation can correlate gbits that are initially uncorrelated; hence not even classical computation is possible.
\end{theorem}
We will now prove this result for the case $d\geq 4$. The proof in the $d=2$ case uses similar techniques, but differs in several details for group-theoretic reasons. It will hence be deferred to the appendix.

As a first step, we will consider the generators of global transformations and show that there exists at least one that is of a certain normal form. This part of the proof is valid for all dimensions $d\geq 2$. A large part of this first step follows the construction in Ref.~\cite{Torre} and extends it to arbitrary dimensions.

\subsection{Generator normal form for all dimensions $d\geq 2$}
\label{SubsecAllD}
\begin{figure}[!hbt]
\includegraphics[angle=0, width=.35\textwidth]{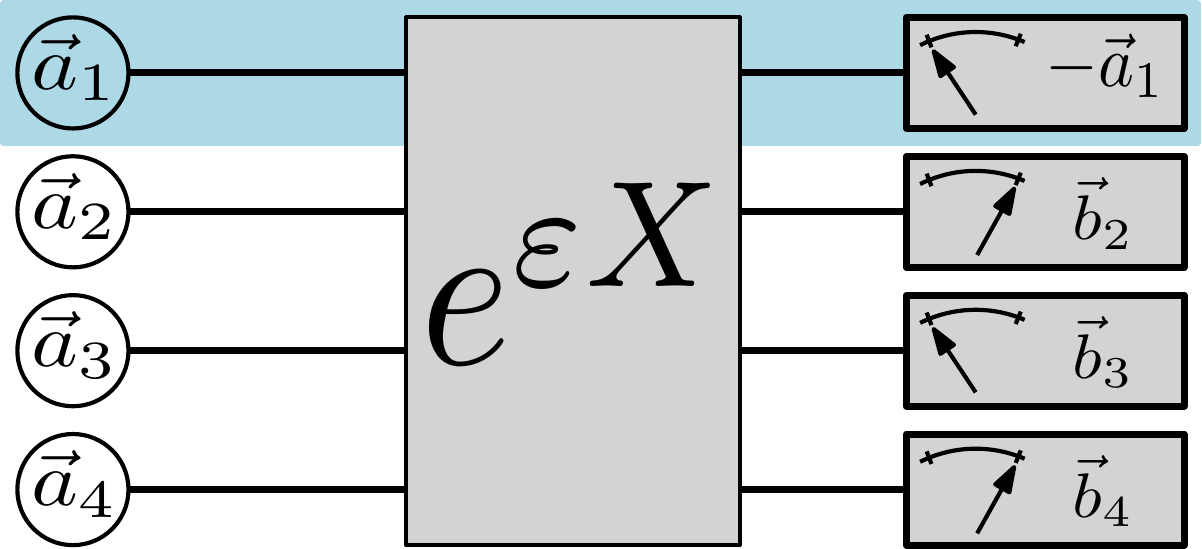}
\caption{We are using configurations like this one to derive constraints on the generators $X\in\mathfrak{g}$. In the special case $\varepsilon=0$, the transformation $\exp(\varepsilon X)$ reduces to the identity. Hence, if we prepare the first wire in the (pure) state with Bloch vector $\vec a_1$, and perform a final measurement of that wire with Bloch vector $-\vec a_1$, the corresponding outcome will have probability zero, regardless of which local measurements we choose for the other wires. But probability zero is a local minimum, which implies that the derivative of this probability with respect to $\varepsilon$ must be zero (yielding $C[\vec a_1]=0$), and the second derivative must be non-negative (yielding constraint~(\ref{eq:2ndOrder1}) in the case $k=1$).}
\label{fig_circuit2}
\end{figure}

Let $G \in \mathcal{G}$ be a transformation of the composite system. Suppose we prepare $n$ gbits initially in states with Bloch vectors $\vec a_1,\ldots,\vec a_n$, evolve the resulting product state via $G$, and perform a final local $n$-gbit measurement with Bloch vectors $\vec b_1,\ldots,\vec b_n$. The probability that the all the $n$ outcomes on the $n$ gbits are ``yes'' is
\[
	2^{-n} v(\vec{b}_1, \vec{b}_2,\ldots, \vec{b}_n)^\top G v(\vec{a}_1, \vec{a}_2,\ldots, \vec{a}_n) \in [0,1].
\]
Let us consider a group element $G=e^{\epsilon X}$ with $X\in\mathfrak{g}$ (the corresponding Lie algebra) and $\varepsilon\in\R$ and expand:
{\small
\[
	v(\vec{b}_1, \ldots, \vec{b}_n)^\top \Big( \id + \epsilon X + \frac{\epsilon^2}{2} X^2 + \mathcal O(\epsilon^3)\Big) v(\vec{a}_1,\ldots, \vec{a}_n) \in [0, 2^n].
\]}
From now on we restrict ourselves to unit length Bloch vectors, i.e.\ $|\vec a_i|=|\vec b_j|=1$ for all $i,j$. We obtain 
\[
	{\mathcal C}[\vec{a}_1] := v(-\vec{a}_1, \vec{b}_2,..., \vec{b}_n)^\top X v(\vec{a}_1, \vec{a}_2, \ldots, \vec{a}_n) = 0
\]
since the zeroth order is zero which is a local minimum as a function of $\epsilon$ (see Figure~\ref{fig_circuit2} for further explanation). Thus the second order contribution has to be non-negative:
\[
	v(-\vec{a}_1, \vec{b}_2,\ldots, \vec{b}_n)^\top X^2 v(\vec{a}_1, \vec{a}_2,\ldots, \vec{a}_n) \ge 0,
\]
or more generally with the roles of qubits $1$ and $k$ exchanged,
\begin{equation}
	v(\vec{b}_1,\ldots,\vec{b}_{k-1},-\vec{a}_k,\vec{b}_{k+1},\ldots \vec{b}_n)^\top X^2 v(\vec{a}_1,\ldots, \vec{a}_n) \ge 0 \label{eq:2ndOrder1}.
\end{equation}
Other first and second order constraints are 
\begin{eqnarray}
v(\vec{a}_1, \vec{a}_2,\ldots, \vec{a}_n)^\top X v(\vec{a}_1, \vec{a}_2,\ldots, \vec{a}_n) &=& 0, \label{eqZeroFirstOrder} \\
v(\vec{a}_1, \vec{a}_2,\ldots, \vec{a}_n)^\top X^2 v(\vec{a}_1, \vec{a}_2,\ldots, \vec{a}_n) &\le& 0 \label{eq:2ndOrder2}
\end{eqnarray}
for analogous reasons as above (since $\vec b_j=\vec a_j$ for all $j$ yields probability one for $\epsilon=0$, which is the global and thus a local maximum). For fixed Bloch vectors $\vec a_2,\ldots,\vec a_n,\vec b_2,\ldots,\vec b_n$, define $W^\alpha_\beta$ as
{\small
	\begin{equation}
	\left[ \vec e_\beta \otimes \begin{pmatrix} 1 \\ \vec b_2 \end{pmatrix} \otimes \ldots \otimes \begin{pmatrix} 1 \\ \vec b_n \end{pmatrix}\right]^\top X \left[ \vec e_\alpha \otimes \begin{pmatrix} 1 \\ \vec a_2 \end{pmatrix} \otimes \ldots \otimes \begin{pmatrix} 1 \\ \vec a_n \end{pmatrix}\right].
	\label{DefW}
	\end{equation}}
The equation $\mathcal{C}[\vec e_i]=0$ implies $W_0^0 + W_0^i - W_i^0 - W_i^i=0$, and $\mathcal{C}[-\vec e_i]=0$ implies $W_0^0 - W_0^i + W_i^0 - W_i^i =0$. Thus, $W_i^i=W_0^0$ and $W_0^i=W_i^0$ for all $i\geq 1$. Since the vectors $\begin{pmatrix} 1 \\ \vec{a} \end{pmatrix}$ linearly span all of $\R^{d+1}$, we get
\begin{align}
	X^{i\ \alpha_2\ \ldots \ \alpha_n}_{i \ \beta_2\ \ldots\ \beta_n} &= X^{0\ \alpha_2\ \ldots\ \alpha_n}_{0\ \beta_2\ \ldots\ \beta_n},\label{eq:Diag}\\
	X^{i\ \alpha_2\ \ldots\ \alpha_n}_{0\ \beta_2\ \ldots\ \beta_n} &= X^{0\ \alpha_2\ \ldots\ \alpha_n}_{i\ \beta_2\ \ldots\ \beta_n}\label{eq:0Symm}
\end{align}
for all $i\geq 1$ and all $\alpha_2,\ldots,\alpha_n,\beta_2,\ldots,\beta_n\geq 0$. Similarly, $\mathcal C[\frac{1}{\sqrt{2}}(\vec e_i + \vec e_j)] = 0$ for $i\ne j$, $i,j \geq 1$ yields
\begin{eqnarray*}
   W_0^0 + \frac 1 {\sqrt{2}} W_0^i &+&\frac 1 {\sqrt{2}} W_0^j -\frac 1 {\sqrt{2}} W_i^0 -\frac 1 2 W_i^i\\
    &&-\frac 1 2 W_i^j - \frac 1 {\sqrt{2}}W_j^0 - \frac 1 2 W_j^i - \frac 1 2 W_j^j=0.
\end{eqnarray*}
Using the results on $W_i^i$ and $W_i^0$ further above, this reduces to $-\frac 1 2 W_i^j - \frac 1 2 W_j^i=0$, and thus
\begin{align}
	X^{i\ \alpha_2\ \ldots \ \alpha_n}_{j \ \beta_2\ \ldots\ \beta_n} &= -X^{j\ \alpha_2\ \ldots\ \alpha_n}_{i\ \beta_2\ \ldots\ \beta_n} \label{eq:Antisymm}
\end{align}
for all $i,j\geq 1$ and $\alpha_2,\ldots,\alpha_n,\beta_2,\ldots,\beta_n\geq 0$. While we have derived~(\ref{eq:Diag}), (\ref{eq:0Symm}) and~(\ref{eq:Antisymm}) for the first gbit, analogous equations hold for all other gbits with labels $2,\ldots,n$.

Let us denote by $\mathcal{A}$ the antisymmetric $(d+1)\times(d+1)$-matrices of the form
\[
\mathcal{A}:=\left\{A_{\bar A}=\left.
	\begin{pmatrix}
	 0\ \vline &  \vec{0}^\top  \\ \hline
	\vec{0}\ \vline & \bar A  \\
	\end{pmatrix} \,\, \right| \,\, \bar A^\top = -\bar A
	\right\},
\]
and by $\mathcal{B}$ the symmetric $(d+1)\times(d+1)$-matrices of the form
\[
\mathcal{B}:=\left\{\left.
	B_{\vec{b}} =
	\begin{pmatrix}
		0 \ \vline & \vec{b}^\top \\ \hline
	\vec b \ \vline & 0 
	\end{pmatrix} \,\,\right|\,\, \vec b \in \R^d\right\}.
\]
Furthermore, let $\mathcal{I}:=\R\cdot \id$, i.e.\ all multiples of the $(d+1)\times(d+1)$ identity matrix. The sets $\mathcal{A}$, $\mathcal{B}$ and $\mathcal{I}$ are real linear matrix subspaces. Note that these three spaces are pairwise orthogonal with respect to the Hilbert-Schmidt inner product $\langle X,Y\rangle:=\tr(X^\top Y)$. The matrix $W$ defined in~(\ref{DefW}) must then be an element of $\mathcal{A}\oplus\mathcal{B}\oplus\mathcal{I}$ due to the identities for its components that we have derived above. More generally, since the same identities hold for every index $i\in\{1,\ldots,n\}$ for the tensor $X$, we obtain $X\in(\mathcal{A}\oplus\mathcal{B}\oplus\mathcal{I})^{\otimes n}$. Since $X\in\mathfrak{g}$ was arbitrary, this tells us that
\[
   \mathfrak{g} \subset (\mathcal{A}\oplus\mathcal{B}\oplus\mathcal{I})^{\otimes n}.
\]
The Lie algebra of the local transformations is
{\small
\[
	\mathfrak{g}_\mathrm{loc} = \mathcal A \otimes \id \otimes \ldots \otimes \id + \id \otimes \mathcal A \otimes \id \otimes \ldots \otimes \id + \ldots + \id \otimes \id \otimes \ldots \otimes \id \otimes \mathcal A,
\]}
writing ``$+$'' instead of ``$\oplus$'' for readability. We can write the space $(\mathcal{A}\oplus\mathcal{B}\oplus\mathcal{I})^{\otimes n}$ in a somewhat different form. To this end, consider strings of symbols $x\in\{A,B,I\}^n$, for example $x=ABAI$ (if $n=4$), and denote the corresponding tensor product matrix spaces by $S_x$; for this example, $S_x=\mathcal{A}\otimes\mathcal{B}\otimes\mathcal{A}\otimes\mathcal{I}$. Then $S_x\perp S_y$ for $x\neq y$ (with respect to the Hilbert-Schmidt inner product), and
\[
   (\mathcal{A}\oplus\mathcal{B}\oplus\mathcal{I})^{\otimes n}=\bigoplus_{x\in\{A,B,I\}^n} S_x.
\]
Now let $X\in\mathfrak{g}\setminus\mathfrak{g}_{\rm loc}$ be an arbitrary generator which is not in the local Lie algebra (here we explicitly make the assumption that such an $X$ exists). Since $X\neq 0$, there must exist $x$ such that $\Phi_x(X)\neq 0$ for the orthogonal projection $\Phi_x$ into $S_x$, and since $X\not\in\mathfrak{g}_{\rm loc}$, at least one of those $x$ must satisfy
\[
   x\not\in\{AI\ldots I, IAI\ldots I,\quad\ldots,\quad I\ldots IA\}.
\]
Reordering the gbits, we may assume that $x=A^{n_A}B^{n_B}I^{n_I}$, where $n_A+n_B+n_I=n$ and one of the following three cases applies:
\begin{itemize}
	\item[(i)] $n_A=0$,
	\item[(ii)] $n_A=1$ and $n_B\geq 1$,
	\item[(iii)] $n_A\geq 2$.
\end{itemize}
Since $S_x$ has an orthonormal basis of matrices of the form $A_{\bar A_1}\otimes\ldots\otimes A_{\bar A_{n_A}}\otimes B_1\otimes\ldots\otimes B_{n_B}\otimes\id^{\otimes n_I}$, where all $A_{\bar A_i}\in\mathcal{A}$ and $B_i\in\mathcal{B}$, there must exist some matrix $\tilde M_x$ of that form (i.e.\ $\tilde M_x\in S_x$) such that $\langle X,\tilde M_x\rangle\neq 0$. By moving constant scalar factors into the $A$-terms, we may assume that there are unit vectors $\vec b_i$ such that $B_i=B_{\vec b_i}$ for $i=1,\ldots,n_B$. But since $\hat R B_{\vec b}\hat R^\top=B_{R\vec b}$ for all $R\in{\rm SO}(d)$, there are orthogonal matrices $\hat R_i$ such that $R_i \vec b_i=\vec e_1=(1,0,\ldots,0)^\top$ for all $i$, and the local transformation $T:=\id^{\otimes n_A}\otimes \hat R_1 \otimes \ldots \otimes \hat R_{n_B}\otimes\id^{\otimes n_I}$ satisfies
\begin{eqnarray*}
   M'_x&:=&T \tilde M_x  T^{-1} = T \tilde M_x T^\top\\ &=& A_{\bar A_1}\otimes\ldots\otimes A_{\bar A_{n_A}} \otimes B^{\otimes n_B} \otimes \id^{\otimes n_I},
\end{eqnarray*}
where $B:=B_{\vec e_1}$. Set $X':=T X T^{-1}$, then since $T\in\mathcal{G}_{\rm loc}\subset \mathcal{G}$ and since the adjoint action of $\mathcal G_{\mathrm{loc}}$ preserves $\mathfrak{g}_{\mathrm{loc}}$, we have $X'\in \mathfrak{g}\setminus\mathfrak{g}_{\rm loc}$, and $\langle X',M'_x\rangle = \tr(T X T^{-1} T \tilde M_x T^{-1})=\langle X,\tilde M_x\rangle\neq 0$. Similar argumentation allows us to bring the $A_{\bar A_i}$ into a standard form. Since the $d\times d$-matrices $\bar A_i$ are antisymmetric, one can infer from the results in~\cite{Thompson} that there are orthogonal transformations $R_i\in{\rm SO}(d)$ such that
{\small
\begin{eqnarray*}
   R_i \bar A_i R_i^\top &=& \begin{pmatrix}
    0 & \lambda_1^{{(i)}} & & & & & \\
    -\lambda_1^{{(i)}} & 0 & & & & & \\
    & & 0 & \lambda_2^{{(i)}} & & & \\
    & & -\lambda_2^{{(i)}} & 0 & & & \\
    & & & & \ddots & &	\\
    & & & & & 0 & \lambda_{d/2}^{{(i)}} \\
    & & & & & -\lambda_{d/2}^{{(i)}} & 0
 \end{pmatrix}
 (d\mbox{ even}),\\
 && \hskip -1em
 \begin{pmatrix}
  0 & & & & & & &\\
    & 0 & \lambda_1^{{(i)}} & & & & & \\
    & -\lambda_1^{{(i)}} & 0 & & & & & \\
    & & & 0 & \lambda_2^{{(i)}} & & & \\
    & & & -\lambda_2^{{(i)}} & 0 & & & \\
    & & & & & \ddots & &	\\
    & & & & & & 0 & \lambda_{\frac{d-1} 2}^{{(i)}} \\
    & & & & & & -\lambda_{\frac{d-1} 2}^{{(i)}} & 0
 \end{pmatrix}
 (d\mbox{ odd}).
\end{eqnarray*}}
To save space, we will use the following notation in the remainder of the paper, where $\sigma=\begin{pmatrix} 0 & 1 \\ -1 & 0\end{pmatrix}$:
\[
   R_i \bar A_i R_i^\top =\left\{
      \begin{array}{cl}
      	  \lambda_1^{{(i)}} \sigma \oplus \lambda_2^{{(i)}} \sigma \oplus \ldots \oplus \lambda_{d/2}^{{(i)}}\sigma & (d\mbox{ even}),\\
      	  0_{1\times 1}\oplus \lambda_1^{{(i)}} \sigma \oplus \lambda_2^{{(i)}} \sigma \oplus \ldots \oplus \lambda_{\frac{d-1}2}^{{(i)}} \sigma & (d\mbox{ odd}).
      \end{array}
   \right.
\]

Now consider the corresponding $(d+1)\times(d+1)$-matrices $A_{R_i \bar A_i R_i^\top}$, for which we will introduce the following notation. By $A_j$, denote the matrix for which only the $j$-th block is non-zero, with $\lambda_j=1$. That is, for even $d$, we have the $(d+1)\times (d+1)$-matrices
\begin{eqnarray*}
A_1 &=& 0_{1\times 1}\oplus \sigma \oplus 0_{2\times 2}\oplus\ldots\oplus 0_{2\times 2},\\
A_2 &=& 0_{1\times 1}\oplus 0_{2\times 2}\oplus \sigma \oplus 0_{2\times 2}\oplus\ldots\oplus 0_{2\times 2}, \\
&\vdots & \\
A_{d/2} &=& 0_{1\times 1}\oplus 0_{2\times 2}\oplus\ldots\oplus 0_{2\times 2}\oplus \sigma,
\end{eqnarray*}
and for odd $d$, we have an extra initial zero, namely
\begin{eqnarray*}
A_1 &=& 0_{2\times 2}\oplus \sigma \oplus 0_{2\times 2}\oplus\ldots\oplus 0_{2\times 2},\\
A_2 &=& 0_{2\times 2}\oplus 0_{2\times 2}\oplus \sigma \oplus 0_{2\times 2}\oplus\ldots\oplus 0_{2\times 2}, \\
&\vdots & \\
A_{(d-1)/2} &=& 0_{2\times 2}\oplus 0_{2\times 2}\oplus\ldots\oplus 0_{2\times 2}\oplus \sigma.
\end{eqnarray*}
The local transformation $\tilde T:=\hat R_1\otimes\ldots\otimes \hat R_{n_A}\otimes \id^{\otimes n_B}\otimes \id^{\otimes n_I}$ satisfies
{\small
\begin{eqnarray}
   M_x:= \tilde T M'_x \tilde T^{-1} = \tilde T M'_x \tilde T^\top \nonumber\\
   =\left(\sum_j \lambda_j^{(1)} A_j\right)\otimes\ldots\otimes\left(\sum_j \lambda_j^{(n_A)} A_j\right)\otimes B^{\otimes n_B} \otimes\id^{\otimes n_I},
   \label{eqSimpleForm}
\end{eqnarray}}
where the $\lambda_j^{(i)}$ are real numbers. Set $X'':=\tilde T X' \tilde T^{-1}$, then since $\tilde T\in\mathcal{G}_{\rm loc}\subset \mathcal{G}$, we have $X''\in \mathfrak{g}\setminus\mathfrak{g}_{\rm loc}$, and $\langle X'',M_x\rangle = \tr(\tilde T X' \tilde T^{-1} \tilde T M'_x \tilde T^{-1})=\langle X',M'_x\rangle\neq 0$.

In summary, we have shown that if there exist any nonlocal generators at all, then there is one (denoted $X''$) that has non-zero overlap with a matrix $M_x\in S_x$ of the simple form~(\ref{eqSimpleForm}).

Next we will show that this implies that $\mathfrak{g}=\mathfrak{g}_{\rm loc}$ for all Bloch ball dimensions $d\geq 4$.

\subsection{Proof of Theorem~\ref{TheMain} for $d\geq 4$}
We now use Schur's Lemma to construct orthogonal projectors (with respect to the Hilbert-Schmidt inner product) onto the subspaces of $\mathcal{A}\oplus\mathcal{B}\oplus\mathcal{I}$. First, define
\[
   \Phi_I[M]:=\int_{{\rm SO}(d)} \hat R M \hat R^{-1}\, dR\qquad (M\in\mathcal{A}\oplus\mathcal{B}\oplus\mathcal{I}),
\]
then $\Phi_I[M]=0$ for all $M\in\mathcal{A}\oplus \mathcal{B}$ and $\Phi_I[M]=M$ for all $M\in\mathcal{I}$. Since these subspaces are orthogonal with respect to the Hilbert-Schmidt inner product, $\Phi_I$ is the orthogonal projector onto the subspace $\mathcal{I}$ of $\mathcal{A}\oplus\mathcal{B}\oplus\mathcal{I}$ (we are not interested in its action on matrices that are not in the space $\mathcal{A}\oplus\mathcal{B}\oplus\mathcal{I}$).

For $j=1,\ldots,d$, consider the stabilizer subgroup
\[
   \mathcal{G}_j:=\{R\in{\rm SO}(d)\,\,|\,\, R \vec e_j = \vec e_j\},
\]
where $\vec e_j$ denotes the $j$th standard unit vector in $\R^d$. Every $\mathcal{G}_j$ is isomorphic to ${\rm SO}(d-1)$ whose fundamental representation is irreducible (note that this is not true for $d=3$; this causes the crucial difference to Ref.~\cite{Torre}). Set
\[
   \Phi_{\vec e_j}[M]:=\int_{\mathcal{G}_j} \hat R M \hat R^{-1}\, dR \qquad (M\in\mathcal{A}\oplus\mathcal{B}\oplus\mathcal{I}),
\]
then $\Phi_{\vec e_1}[M]=\bigintss_{{\rm SO}(d-1)} \begin{pmatrix} \id_2 & \\ & S\end{pmatrix} M \begin{pmatrix} \id_2 & \\ & S^{-1}\end{pmatrix}\, dS$, and, similarly as above, Schur's Lemma implies that $\Phi_{\vec e_1}$ is the orthogonal projector onto ${\rm span}(B)\oplus \mathcal{I}$. Hence $\Phi_B:=\Phi_{\vec e_1}-\Phi_I$ is the orthogonal projector onto ${\rm span}(B)$.

Finally, we will construct the orthogonal projector onto $\mathcal{A}_{\rm blocks}:={\rm span}\{A_1,\ldots,A_z\}$, where $z=d/2$ if $d$ is even and $z=(d-1)/2$ if $d$ is odd. To this end, define the ${\rm SO}(2)$-matrix $R(\theta):=\begin{pmatrix} \cos\theta & \sin\theta \\ -\sin\theta & \cos\theta \end{pmatrix}$, and set
\[
   \hat R(\theta_1,\theta_2,\ldots,\theta_z):=\begin{pmatrix} \id_y & & & \\
 & R(\theta_1) & & \\ & & \ddots & \\ & & & R(\theta_z)	
\end{pmatrix},
\]
where $y=1$ if $d$ is even and $y=2$ if $d$ is odd. Furthermore, define $\Phi'[M]$ as
{\small
\[
   \int_0^{2\pi} \frac{d\theta_1}{2\pi} \int_0^{2\pi} \frac{d\theta_2}{2\pi} \ldots \int_0^{2\pi} \frac{d\theta_z}{2\pi} \hat R(\theta_1,\ldots,\theta_z) M \hat R(\theta_1,\ldots,\theta_z)^{-1}.
\]}
Using the identities
\begin{widetext}
\[
   \int_0^{2\pi} R(\theta)\,\frac{d\theta}{2\pi}=0,\qquad
   \int_0^{2\pi} R(\theta)\begin{pmatrix}m_{11} & m_{12} \\ m_{21} & m_{22} \end{pmatrix} R(-\theta)\frac{d\theta}{2\pi}=
   \frac 1 2 \begin{pmatrix} m_{11}+m_{22} & m_{12} - m_{21} \\ -m_{12}+m_{21} & m_{11}+ m_{22}\end{pmatrix}=:\Psi\left[\begin{pmatrix}m_{11} & m_{12} \\ m_{21} & m_{22} \end{pmatrix}\right],
\]
\end{widetext}
we can evaluate the action of $\Phi'$ as follows. First, any given $(d+1)\times(d+1)$-matrix $M$ can be written in the block matrix form
\[
   M=\begin{pmatrix}
   	   M_{0,0} & \hdots & M_{0,z} \\
   	   \vdots & \ddots & \vdots \\
   	   M_{z,0} & \hdots & M_{z,z}
   \end{pmatrix}
\]
where $M_{0,0}$ is a $y\times y$-matrix, all $M_{i,j}$ for $i,j\geq 1$ are $2\times 2$-matrices, and the other matrices are $y\times 2$ and $2\times y$-matrices. Then, the action of $\Phi'$ becomes
\[
   \Phi'[M]=\begin{pmatrix}
   	   M_{0,0} & 0 & \hdots & 0 \\
   	   0 & \Psi[M_{1,1}] & & \vdots \\
   	   \vdots & & \ddots & 0 \\ 
   	   0 & \hdots & 0 & \Psi[M_{z,z}]
   \end{pmatrix}.
\]
Hence $\Phi'$ is an orthogonal projection that acts as the identity on $\mathcal{I}$ (i.e.\ $\Phi'(\id)=\id$), and it projects $\mathcal{A}$ into its subspace $\mathcal{A}_{\rm blocks}$. Furthermore, if $d$ is even, then $\Phi'$ annihilates $\mathcal{B}$, and if $d$ is odd, then $\Phi'$ projects $\mathcal{B}$ into its subspace ${\rm span}(B)$. Thus, for $d$ even, the orthogonal projector onto $\mathcal{A}_{\rm blocks}$ is $\Phi_A:=\Phi'-\Phi_I$, and for $d$ odd, it is $\Phi_A:=\Phi'-\Phi_I-\Phi_B$. Note that all these statements are only claimed to hold for the case that the maps are applied to operators in $\mathcal{A}\oplus\mathcal{B}\oplus\mathcal{I}$.

The projectors $\Phi_I$, $\Phi_B$ and $\Phi_A$ map the Lie algebra $\mathfrak{g}$ into itself, if we apply different products of those projectors to the $n$ sites. For example, consider the special case $n=1$. Then $Z\in\mathfrak{g}$ implies $\Phi_I[Z]\in\mathfrak{g}$ since $\mathfrak{g}$ is closed with respect to conjugations by elements of $\mathcal{G}$ and integrals. Similarly, $\Phi_{\vec e_1}[Z]\in\mathfrak{g}$, and since $\mathfrak{g}$ is a linear space, we also have $\Phi_B[Z]=\Phi_{\vec e_1}[Z]-\Phi_I[Z]\in\mathfrak{g}$, and similarly for the projector $\Phi_A$. If $n\geq 2$, then we can successively apply the projectors to one of the sites, using the fact that tensoring local rotations with identities gives local transformations in $\mathcal{G}_{\rm loc}$. Thus, if we define
\[
   \Phi:=\Phi_A^{\otimes n_A}\otimes\Phi_B^{\otimes n_B}\otimes\Phi_I^{\otimes n_I},
\]
then $Y:=\Phi[X'']$ is another valid generator, $Y\in\mathfrak{g}$. Furthermore, $\Phi[M_x]=M_x$, hence
\begin{equation}
   0\neq \langle X'',M_x\rangle = \langle X'',\Phi[M_x]\rangle
   =\langle \Phi[X''],M_x\rangle = \langle Y,M_x\rangle
   \label{eqCalculation}
\end{equation}
and thus $Y\neq 0$ (we have used that $\Phi$ is an orthogonal projection and thus in particular self-adjoint with respect to the Hilbert-Schmidt inner product). In particular, $Y\in{\rm Im}(\Phi) = \mathcal{A}_{\rm blocks}^{\otimes n_A}\otimes {\rm span}(B)^{\otimes n_B}\otimes \mathcal{I}^{\otimes n_I}$. Consequently, there are real numbers $\lambda_{j_1,\ldots,j_{n_A}}$ such that
\[
   Y=\sum_{j_1,\ldots,j_{n_A}=1}^z \lambda_{j_1,\ldots,j_{n_A}} A_{j_1}\otimes \ldots \otimes A_{j_{n_A}} \otimes B^{\otimes n_B} \otimes \id^{\otimes n_I}.
\]
Now we apply the identities $A_j A_k = -\delta_{jk} P_j$ and $B^2=P_B$, where
\begin{eqnarray*}
P_B &=& \mathbf{1}_{2\times 2} \oplus 0_{(d-1)\times(d-1)},\\
P_1&=&0_{y\times y}\oplus \mathbf{1}_{2\times 2}\oplus 0_{2(z-1)\times 2(z-1)},\\
P_2&=& 0_{y\times y} \oplus 0_{2\times 2}\oplus \mathbf{1}_{2\times 2}\oplus 0_{2(z-2)\times 2(z-2)}
\end{eqnarray*}
and so on, up to $P_z$. This gives us
{\small
\begin{equation}
   Y^2 = (-1)^{n_A} \sum_{j_1,\ldots,j_{n_A}} \lambda_{j_1,\ldots,j_{n_A}}^2 P_{j_1}\otimes \ldots \otimes P_{j_{n_A}} \otimes P_B^{\otimes n_B} \otimes \id^{\otimes n_I}.
   \label{eqYsquared}
\end{equation}}
Suppose that $n_A$ is even so that $(-1)^{n_A}=1$. We will now show that constraint~(\ref{eq:2ndOrder2}) gets violated. To this end, fix some $j_1^0,\ldots,j_{n_A}^0$ such that $\lambda_{j_1^0,\ldots,j_{n_A}^0}\neq 0$. For $i=1,\ldots,n_A$, choose some unit vector $\vec a_i\in\R^d$ such that $\begin{pmatrix}1 \\ \vec a_i\end{pmatrix}^\top P_{j_i^0} \begin{pmatrix}1 \\ \vec a_i\end{pmatrix}>0$; for all other $j_i$, we automatically get $\begin{pmatrix}1 \\ \vec a_i\end{pmatrix}^\top P_{j_i} \begin{pmatrix}1 \\ \vec a_i\end{pmatrix}\geq 0$. For $i=n_A+1,\ldots,n_A+n_B$, set $\vec a_i:=\vec e_1$, then $\begin{pmatrix}1 \\ \vec a_i\end{pmatrix}^\top P_B \begin{pmatrix}1 \\ \vec a_i\end{pmatrix}=2$. Finally, for $i\geq n_A+n_B+1$, choose $\vec a_i$ arbitrarily such that $\begin{pmatrix}1 \\ \vec a_i\end{pmatrix}^\top \id \begin{pmatrix}1 \\ \vec a_i\end{pmatrix}=2$. Altogether, we obtain
\[
   v(\vec a_1,\ldots,\vec a_n)^\top\, Y^2\,v(\vec a_1,\ldots,\vec a_n)>0
\]
which violates constraint~(\ref{eq:2ndOrder2}). Thus $n_A$ must be odd, and $(-1)^{n_A}=-1$.

Recall constraint~(\ref{eq:2ndOrder1}) in the special case $k=2$:
\begin{equation}
   v(\vec b_1,-\vec a_2,\vec b_3,\ldots,\vec b_n)^\top \, Y^2 \, v(\vec a_1,\vec a_2,\ldots,\vec a_n)\geq 0
   \label{eqConstraintNew}
\end{equation}
for all unit vectors $\vec a_i, \vec b_j\in\R^d$. For all $i \in [n_A+n_B+1,n]\setminus\{2\}$, choose $\vec a_i,\vec b_i$ such that $\begin{pmatrix}1 \\ \vec b_i\end{pmatrix}^\top \id \begin{pmatrix}1 \\ \vec a_i\end{pmatrix}>0$ (simply avoid the choice $\vec a_i=-\vec b_i$). Similarly, for all $i\in[n_A+1,n_A+n_B]\setminus\{2\}$, choose $\vec a_i,\vec b_i$ such that $\begin{pmatrix}1 \\ \vec b_i\end{pmatrix}^\top P_B \begin{pmatrix}1 \\ \vec a_i\end{pmatrix}>0$. We will now distinguish two cases for $n_A$.

First, consider the case $n_A=1$. Since our original generator $X$ was chosen nonlocal, it follows that $n_B\geq 1$, as explained in Subsection~\ref{SubsecAllD}. Thus, the second tensor factor in~(\ref{eqYsquared}) must be $P_B$. We will now choose $\vec a_2=\vec e_2$ which implies that $\begin{pmatrix}1 \\ -\vec a_2\end{pmatrix}^\top P_B \begin{pmatrix}1 \\ \vec a_2\end{pmatrix}=1$. But then we may still choose $\vec b_1,\vec a_1$ arbitrarily, and by choosing these two unit vectors suitably from the subspace ${\rm Im}(P_{j_1})$, we may generate an arbitrary sign for $\begin{pmatrix}1 \\ \vec b_1\end{pmatrix}^\top P_{j_1} \begin{pmatrix}1 \\ \vec a_1\end{pmatrix}$. Thus, we can break constraint~(\ref{eqConstraintNew}) by a suitable choice of these two unit vectors, which yields a contradiction.

Second, suppose that $n_A\geq 3$ (we already know that $n_A$ must be odd). Then we can choose $\vec a_2$ such that $\begin{pmatrix}1 \\ -\vec a_2\end{pmatrix}^\top P_{j_2} \begin{pmatrix}1 \\ \vec a_2\end{pmatrix}=-1$. We have even more freedom than in the previous case: for all $i\in [1,n_A]\setminus\{2\}$, we can choose $\vec b_i,\vec a_i$ from the subspace ${\rm Im}(P_{i_j})$ such that we get an arbitrary sign for every $\begin{pmatrix}1 \\ \vec b_i\end{pmatrix}^\top P_{j_i} \begin{pmatrix}1 \\ \vec a_i\end{pmatrix}$. This also leads to a violation of constraint~(\ref{eqConstraintNew}), and we obtain a contradiction as well.

This means that our initial assumption must have been wrong --- namely, that there exists a generator in $\mathfrak{g}\setminus\mathfrak{g}_{\rm loc}$. We conclude that instead this set must be empty, hence $\mathfrak{g}=\mathfrak{g}_{\rm loc}$. But since $\mathcal{G}$ is compact and connected, it follows from~\cite[Theorem VII.2.2 (v)]{Simon} that $\mathcal{G}$ cannot be larger than $\mathcal{G}_{\rm loc}$. This proves our main result, Theorem~\ref{TheMain}, for Bloch ball dimensions $d\geq 4$. The proof for $d=2$ is given in Appendix~\ref{AppendixProof}.

\section{Conclusions}
\label{SecConclusions}
Given a few simple properties that turn out to characterize qubit quantum computation, we have considered a natural modification: allowing the single bits to have more or less than the qubit's $d=3$ degrees of freedom. We have analyzed the set of possible reversible transformations in the resulting theories, under the conjecture~\cite{DakicBrukner} (and in hopes) of discovering novel computational models that differ in interesting ways from quantum computation. Unfortunately, it turns out that the resulting models do not allow for any non-trivial reversible gates whatsoever. This reinforces earlier intuition~\cite{Aaronson} that quantum theory, or in this context quantum computation, is an ``island in theoryspace''.

While we have made an effort to be as careful and parsimonious in our assumptions as possible, it is still interesting to ask whether there are any remaining ``loopholes'' that could in principle leave some wiggle room for non-trivial beyond-quantum computation: can any of the assumptions of Subsection~\ref{SubsecSeveralGbits} be dropped or weakened, while insisting that single bits are described by Bloch balls? We discuss several options in Appendix~\ref{SecAssumptions}; in short, the most promising (but difficult) approaches would be to drop tomographic locality, and/or to drop reversibility or continuity of transformations. Both options present formidable mathematical challenges and are therefore deferred to future work.

The ``rigidity'' of quantum theory, i.e.\ the difficulty of modifying it in consistent ways, has been recognized in different contexts for a long time, see e.g.\ Weinberg's proposal of a nonlinear modification of quantum mechanics~\cite{Weinberg}, and Gisin's subsequent discovery~\cite{Gisin} that this modification allows for superluminal signalling. The research presented in this paper and in other work (like~\cite{Galley1,Galley2}) makes this intuition more rigorous by specifying which combinations of principles already enforce the familiar behavior of quantum theory. These insights also illuminate our understanding of quantum computation, since they tell us which physical principles enforce its properties, and/or which other theoretical models of computation are plausibly conceivable.

Finally, it is interesting to speculate that the result of  this paper is indirectly related to spacetime physics. After all, it is the fact that a qubit is represented as a 3-ball $B^3$, with ${\rm SO}(3)$ as its transformation group, which allows for spin-$1/2$ particles that couple to rotations in three-dimensional space. Given the popularity of approaches in which spacetime emerges in some way from an underlying quantum theory~\cite{Jacobson,Ryu,HoehnEssay}, this observation can perhaps be regarded as more than a coincidence. In fact, it has been argued more rigorously that the structures of quantum theory and spacetime mutually constrain each other~\cite{MM3D,DakicBrukner,HoehnMueller,Linearity,Garner}. This suggests a slogan that also fits some other ideas from quantum information~\cite{AaronsonNP}: the limits of computation are the limits of our world.

\bigskip

\textbf{Acknowledgments.} We thank \v{C}aslav Brukner, Borivoje Daki\'c, Philip Goyal, Llu\'is Masanes, and Jochen Rau for discussions. M.K.\ acknowledges support from the Austrian Science Fund (FWF) through the Doctoral Programme CoQuS (W1210). This publication was made possible through the support of a grant from the John Templeton Foundation. The opinions expressed in this publication are those of the authors and do not necessarily reflect the views of the John Templeton Foundation. This research was supported in part by Perimeter Institute for Theoretical Physics. Research at Perimeter Institute is supported by the Government of Canada through the Department of Innovation, Science and Economic Development Canada and by the Province of Ontario through the Ministry of Research, Innovation and Science.

\onecolumngrid

\appendix
\section{Proof of Theorem~\ref{TheMain} for $d=2$}
\label{AppendixProof}

Due to different group-theoretic properties, we now have less freedom to construct projectors by integrating over conjugations with local transformations. A first difference to the case $d\geq 4$ appears already in
\[
   \Phi_{AI}[M]:=\int_{{\rm SO}(2)} \hat R M \hat R^{-1}\, dR \qquad (M\in\mathcal{A}\oplus\mathcal{B}\oplus\mathcal{I}).
\]
It turns out that this map leaves not only $\mathcal{I}$ but also $\mathcal{A}$ (which is now one-dimensional) invariant. Since it still annihilates $\mathcal{B}$, it is the orthogonal projector onto $\mathcal{A}\oplus\mathcal{I}$. We can still use $\Phi_B:=\id-\Phi_{AI}$ as the projector onto $\mathcal{B}$, but we cannot construct a projector onto ${\rm span}(B)$ in a similar way. Now set $n_{AI}:=n_A+n_I$, and reorder the gbits such that $A$ comes first, and then $I$, and then $B$ (in contrast to the previous subsections). Next define the orthogonal projector
\[
   \Phi:=\Phi_{AI}^{\otimes n_{AI}}\otimes\Phi_B^{\otimes n_B},
\]
then $Y:=\Phi[X'']$ is another valid generator, i.e.\ $Y\in\mathfrak{g}$, and $Y\in (\mathcal{A}\oplus\mathcal{I})^{\otimes n_{AI}}\otimes\mathcal{B}^{\otimes n_B}$. Since $\Phi[M_x]=M_x$, the calculation~(\ref{eqCalculation}) proves that $Y\neq 0$. It also follows that $Y\in\mathfrak{g}\setminus\mathfrak{g}_{\rm loc}$ since $Y$ has non-zero overlap with $M_x$ which in turn is orthogonal onto $\mathfrak{g}_{\rm loc}$. Defining $A^{(0)}:=\id$ and $A^{(1)}:=\begin{pmatrix} 0 & 0 & 0 \\ 0 & 0 & 1 \\ 0 & -1 & 0 \end{pmatrix}$ (which spans the one-dimensional space $\mathcal{A}$), $B_0:=\begin{pmatrix} 0 & 1 & 0 \\ 1 & 0 & 0 \\ 0 & 0 & 0 \end{pmatrix}$ and $B_1:=\begin{pmatrix} 0 & 0 & 1 \\ 0 & 0 & 0 \\ 1 & 0 & 0 \end{pmatrix}$, the generator $Y$ can be written in the form
\begin{equation}
   Y=\sum_{k_1,\ldots,k_n=0}^1 \alpha_{k_1,\ldots,k_n} A^{(k_1)}\otimes A^{(k_2)}\otimes\ldots\otimes A^{(k_{n_{AI}})}\otimes B_{k_m}\otimes \ldots \otimes B_{k_n},
   \label{eqFormOfY}
\end{equation}
where the $\alpha_{k_1,\ldots,k_n}$ are real numbers and $m:=n_{AI}+1$.

Now we will apply the first-order constraint~(\ref{eqZeroFirstOrder}) for some special choice of unit vectors $\vec a_i$. First, fix $j_1,j_2,\ldots,j_n\in\{0,1\}$ arbitrarily. For $i\leq n_{AI}$ set $\vec a_i:=\vec e_1$, and for $i\geq m$ set
\[
   \vec a_i:=\left\{
      \begin{array}{cl}
      	  \vec e_1 & \mbox{if }j_i=0 \\
      	  \vec e_2 & \mbox{if }j_i=1.
      \end{array}
   \right.
\]
We obtain the following two equations
\[
\begin{pmatrix}1 \\ \vec a_i\end{pmatrix}^\top A^{(k_i)} \begin{pmatrix}1 \\ \vec a_i\end{pmatrix}=2\delta_{k_i,0}\quad (i=1,\ldots,n_{AI}),\qquad
\begin{pmatrix}1 \\ \vec a_i\end{pmatrix}^\top B_{k_i} \begin{pmatrix}1 \\ \vec a_i\end{pmatrix}=2\delta_{j_i,k_i}\quad (i=m,\ldots,n),
\]
and substituting them into constraint~(\ref{eqZeroFirstOrder}) yields
\[
   0=v(\vec a_1,\ldots,\vec a_n)^\top Y v(\vec a_1,\ldots,\vec a_n)=2^{n_{AI}}\sum_{k_m,\ldots,k_n=0}^1 \alpha_{0,\ldots,0,k_m,\ldots,k_n} \prod_{\ell=m}^n \begin{pmatrix}1 \\ \vec a_{\ell
   }\end{pmatrix}^\top B_{k_{\ell}} \begin{pmatrix}1 \\ \vec a_{\ell}\end{pmatrix}
   =2^n \alpha_{0,\ldots,0,j_m,\ldots,j_n}.
\]
Thus $\alpha_{0,\ldots,0,j_m,\ldots,j_n}=0$, i.e.\ every non-vanishing summand in~(\ref{eqFormOfY}) contains at least one $A^{(1)}$-term. Furthermore, in the special case that $n_B=0$, all summands with a single $A^{(1)}$-term are themselves elements of $\mathfrak{g}_{\rm loc}$, and by subtracting those elements, we obtain another non-zero generator (which now also call $Y$) for which every non-vanishing summand has at least \emph{two} $A^{(1)}$-terms.

Next we slightly generalize constraint~(\ref{eq:2ndOrder1}):

\begin{lemma}
The constraint
\begin{equation}
   v(\vec b_1,\vec b_2,\ldots,\vec b_{k-1},-\vec a_k,\vec b_{k+1},\ldots,\vec b_n)^\top X^2 v(\vec a_1,\vec a_2,\ldots,\vec a_n)\geq 0
   \label{eqConstraintZeros}
\end{equation}
also holds if we replace one or more of the unit vectors $\vec b_j,\vec a_j$, but not $\pm \vec a_k$, by the zero vector.
\end{lemma}
\proof
We start with constraint~(\ref{eq:2ndOrder1}), where all vectors are assumed to be unit vectors. To replace, for example, $\vec b_j$ (for $j\neq k$) by $\vec 0$, consider~(\ref{eq:2ndOrder1}) and its version with $\vec b_j$ replaced by $-\vec b_j$. Adding up the two inequalities (and dividing the result by two) proves~(\ref{eqConstraintZeros}) for $\vec b_j=0$. We can similarly replace any of the $\vec a_j$ (for $j\neq k$) by $\vec 0$, and do so recursively.
\qed

Now we are ready to state and prove the main result of the appendix:
\begin{lemma}
If $d=2$ then $\mathcal{G}=\mathcal{G}_{\rm loc}$, i.e.\ the only reversible transformations are the local transformations.
\end{lemma}
\proof
Our strategy is to prove the following claim:

\underline{Claim:} Let $0\leq\ell\leq n_{AI}$ be an integer. Then $Y$ does not contain any summand in~(\ref{eqFormOfY}) which has exactly $\ell$ occurrences of $A^{(0)}$. In more formal words, if $j_1,\ldots,j_n$ has the property that $\#\{i\in [1,n_{AI}]\,\,|\,\, j_i=0\}=\ell$ then $\alpha_{j_1,\ldots,j_n}=0$.

This claim will then imply that $Y=0$, which is a contradiction (we have shown further above that $Y\neq 0$).

We will prove this claim for two different cases separately; in both cases, our proof will be by induction. Note that we have already shown the claim above for $\ell=n_{AI}$ (since there must be at least one $A^{(1)}$-term in every summand).

\textbf{Case 1:} $n_B=0$ (such that $n_{AI}=n$).

\underline{Induction start:} We know the claim is true for $\ell=n$. Furthermore, since $n_B=0$, we have constructed $Y$ such that no summand contains exactly one $A^{(1)}$-term, hence the claim is also true for $\ell=n-1$.

\underline{Induction hypothesis:} Consider an arbitrary integer $\ell$ with $0\leq \ell\leq n-2$. Let us assume that for any integer $\ell'$ with $0\leq \ell'\leq n$ and $\ell'>\ell$ we know that $Y$ contains no summand with exactly $\ell'$ occurrences of $A^{(0)}$.

\underline{Induction step:} Using the induction hypothesis, we will now show that the Claim also holds for $\ell$ itself.

We do so by contradiction. Suppose there was at least one non-vanishing summand in $Y$ with exactly $\ell$ occurrences of $A^{(0)}$. That is, there exist $j_1^0,\ldots,j_n^0$ such that $\alpha_{j_1^0,\ldots,j_n^0}\neq 0$ and exactly $\ell$ of the $j_i^0$ are equal to zero. We will apply constraint~(\ref{eqConstraintZeros}) for some choice of vectors $\vec a_i,\vec b_i$. To this end, for every $i$ with $j_i^0=0$ set $\vec a_i:=\vec 0$. For those $i$, it follows that $A^{(j_i)} A^{(k_i)} \begin{pmatrix} 1 \\ \vec a_i\end{pmatrix} = \delta_{j_i,0} \delta_{k_i,0} \begin{pmatrix} 1 \\ 0 \\ 0 \end{pmatrix}$. Now $Y^2$ is of the form
\[
   Y^2=\sum_{j_1,\ldots,j_n=0}^1 \sum_{k_1,\ldots,k_n=0}^1 \alpha_{j_1,\ldots,j_n}\alpha_{k_1,\ldots,k_n} (A^{(j_1)}A^{(k_1)})\otimes\ldots \otimes (A^{(j_n)}A^{(k_n)}).
\]
Now consider $w:=Y^2 v(\vec a_1,\ldots,\vec a_n)$. If a summand of $Y^2$ has less than $\ell$ indices $k_i$ with $k_i=0$ then it does not contribute to $w$; also, there are no summands with more than $\ell$ indices $k_i$ with $k_i=0$. Among those summands with exactly $\ell$ indices $k_i$ with $k_i=0$, these indices must occur in exactly those places $i$ where $j_i^0=0$, otherwise those summands do not contribute to $w$. But this enforces that only the summand with $(k_1,\ldots,k_n)=(j_1,\ldots,j_n)=(j_1^0,\ldots,j_n^0)$ contributes to $w$, and we get
\[
   Y^2 v(\vec a_1,\ldots,\vec a_n)=\alpha_{j_1^0,\ldots,j_n^0}^2 \bigotimes_{z=1}^n \left(A^{(j_z^0)}\right)^2 \begin{pmatrix} 1 \\ \vec a_z \end{pmatrix}.
\]
There are at least two indices $z$ with $j_z^0=1$; let $k$ be one of those indices, and define $\vec a_k:=\vec e_1$. Then $\begin{pmatrix} 1 \\ -\vec a_k\end{pmatrix}^\top \left(A^{(j_k^0)}\right)^2 \begin{pmatrix} 1 \\ \vec a_k\end{pmatrix}=1$. Among the remaining places $z$ with $j_z^0=1$, we can choose $\vec a_z$ and $\vec b_z$ such that $\begin{pmatrix} 1 \\ \vec b_z\end{pmatrix}^\top \left(A^{(j_z^0)}\right)^2 \begin{pmatrix} 1 \\ \vec a_z\end{pmatrix}$ takes any sign we like. This will allow is to violate constraint~(\ref{eqConstraintZeros}), and we have a contradiction.

\textbf{Case 2:} $n_B\geq 1$.

\underline{Induction start:} We have already shown the claim for $\ell=n_{AI}$.

\underline{Induction hypothesis:} Consider an arbitrary integer $\ell$ with $0\leq \ell\leq n_{AI}-1$. Let us assume that for any integer $\ell'$ with $0\leq \ell'\leq n_{AI}$ and $\ell'>\ell$ we know that $Y$ contains no summand with exactly $\ell'$ occurrences of $A^{(0)}$.

\underline{Induction step:} We proceed similarly as in Case 1. Using the induction hypothesis, we will now show that the Claim also holds for $\ell$ itself.

We do so by contradiction. Suppose there was at least one non-vanishing summand in $Y$ with exactly $\ell$ occurrences of $A^{(0)}$. That is, there exist $j_1^0,\ldots,j_n^0$ such that $\alpha_{j_1^0,\ldots,j_n^0}\neq 0$ and exactly $\ell$ of the $j_i^0$ among $i\in [1,n_{AI}]$ are equal to zero. We will apply constraint~(\ref{eqConstraintZeros}) for some choice of vectors $\vec a_i,\vec b_i$. To this end, for every $i$ with $j_i^0=0$ set $\vec a_i:=\vec 0$ and choose $\vec b_i$ arbitrarily. For those $i$, it follows that
\begin{equation}
\begin{pmatrix} 1 \\ \vec b_i \end{pmatrix}^\top A^{(j_i)} A^{(k_i)} \begin{pmatrix} 1 \\ \vec a_i\end{pmatrix} = \delta_{j_i,0} \delta_{k_i,0}.
\label{eqAAAA}
\end{equation}
In Case 2, $Y^2$ is of the form
\[
   Y^2=\sum_{j_1,\ldots,j_n=0}^1 \sum_{k_1,\ldots,k_n=0}^1 \alpha_{j_1,\ldots,j_n}\alpha_{k_1,\ldots,k_n} (A^{(j_1)}A^{(k_1)})\otimes\ldots \otimes (A^{(j_{n_{AI}})}A^{(k_{n_{AI}})})\otimes (B_{j_m} B_{k_m})\otimes\ldots\otimes (B_{j_n}B_{k_n}).
\]
Again, we have to choose which place corresponds to the $k$ in constraint~(\ref{eqConstraintZeros}). This time, we will choose $k=m$, and set $\vec a_k=\vec e_1$ if $j_k^0=1$ resp.\ $\vec a_k=\vec e_2$ if $j_k^0=0$, which implies
\[
   \begin{pmatrix} 1 \\ -\vec a_k \end{pmatrix}^\top B_{j_k} B_{k_k} \begin{pmatrix} 1 \\ \vec a_k\end{pmatrix} = \delta_{j_k,j_k^0} \delta_{k_k,j_k^0}.
\]
For all other $i\in [m,n]\setminus\{k\}$ we make the following choice. If $j_i^0=1$ we set $\vec b_i=-\vec e_1$ and $\vec a_i=\vec e_1$, and if $j_i^0=0$ we set $\vec b_i=-\vec e_2$ and $\vec a_i=\vec e_2$. This enforces
\[
   \begin{pmatrix} 1 \\ \vec b_i \end{pmatrix}^\top B_{j_i} B_{k_i} \begin{pmatrix} 1 \\ \vec a_i\end{pmatrix} = \delta_{j_i,j_i^0} \delta_{k_i,j_i^0}\qquad (i\in[m,n]\setminus\{k\}).
\]
Regardless of how we choose the remaining $\vec a_i$, we obtain
\begin{eqnarray}
v(\vec b_1,\vec b_2,\ldots,\vec b_{k-1},-\vec a_k,\vec b_{k+1},\ldots,\vec b_n)^\top Y^2 v(\vec a_1,\vec a_2,\ldots,\vec a_n)=\nonumber \\
   = \sum_{j_1,\ldots,j_n=0}^1 \sum_{k_1,\ldots,k_n=0}^1 \alpha_{j_1,\ldots,.j_n} \alpha_{k_1,\ldots,k_n} \prod_{z=0}^{n_{AI}} \left[ \begin{pmatrix} 1 \\ \vec b_{z}\end{pmatrix}^\top A^{(j_{z})} A^{(k_{z})}\begin{pmatrix} 1 \\ \vec a_{z}\end{pmatrix}\right] \prod_{z\geq m, z\neq k} \left[ \begin{pmatrix} 1 \\ \vec b_{z}\end{pmatrix}^\top B_{j_{z}} B_{k_{z}}\begin{pmatrix} 1 \\ \vec a_{z}\end{pmatrix}\right]\times \nonumber\\
   \times \left[\begin{pmatrix} 1 \\ -\vec a_k \end{pmatrix}^\top B_{j_k} B_{k_k} \begin{pmatrix} 1 \\ \vec a_k\end{pmatrix}\right]\nonumber \\
   = \sum_{j_1,\ldots,j_{n_A}=0}^1 \sum_{k_1,\ldots,k_{n_A}=0}^1 \alpha_{j_1,\ldots,j_{n_A},j_m^0,\ldots,j_n^0} \alpha_{k_1,\ldots,k_{n_A},j_m^0,\ldots,j_n^0} \prod_{z=0}^{n_{AI}}\left[ \begin{pmatrix} 1 \\ \vec b_{z}\end{pmatrix}^\top A^{(j_{z})} A^{(k_{z})}\begin{pmatrix} 1 \\ \vec a_{z}\end{pmatrix}\right].
   \label{eqComplicatedLikeHell}
\end{eqnarray}
Consider the different possibilities for $k_1,\ldots,k_n$ for which $\alpha_{k_1,\ldots,k_{n_{AI}},j_m^0,\ldots,j_n^0}\neq 0$. There are less than or equal to $\ell$ many occurrences of $k_i$ $(1\leq i \leq n_{AI})$ with $k_i=0$. If there are less, then~(\ref{eqAAAA}) implies that the final product in~(\ref{eqComplicatedLikeHell}) vanishes, hence the corresponding summand does not contribute to~(\ref{eqComplicatedLikeHell}). On the other hand, if there are exactly $\ell$ many, then~(\ref{eqAAAA}) implies that this product vanishes unless the occurrences of $k_i=0$ agree with the occurrences of $j_i^0=0$. Similar argumentation works for the $j_1,\ldots,j_n$, and if we also use~(\ref{eqConstraintZeros}), we finally get
\[
   0 \leq \underbrace{\alpha_{j_1^0,\ldots,j_n^0}^2}_{\neq 0} \prod_{1\leq z\leq n_{AI}:j_{z}^0 =1} \left[ \begin{pmatrix} 1 \\ \vec b_{z}\end{pmatrix}^\top \left(A^{(1)}\right)^2 \begin{pmatrix} 1 \\ \vec a_{z}\end{pmatrix}\right].
\]
The product runs over $n_{AI}-\ell$ many indices, so there is at least one $z$ such that $1\leq z \leq n_{AI}$ and $j_z^0=1$. We have not yet chosen the corresponding $\vec b_z$ and $\vec a_z$; it is easy to see that we can choose them so that the terms in the product in the previous expression attain any sign we want. This produces a contradiction, like in Case 1.
\qed

\section{Which assumptions could possibly be dropped or weakened?}
\label{SecAssumptions}

 One candidate assumption that one might consider to weaken is the assumption that the group of single-gbit reversible transformations must be ${\rm SO}(d)$. It is natural to assume that this group must be able to map every pure gbit state to any other (and thus be transitive on the $(d-1)$-sphere). In fact, for odd $d \ne 7$, this demand already singles out $\mathrm{SO}(d)$. However, if $d$ is even or $d=7$, then there are other transitive groups (such as ${\rm SU}(2)$ for $d=4$), and the analysis of the present paper is in principle applicable to this more general situation. The case of $n=2$ gbits has been treated in this more general setting in~\cite{MMAPG2}. There it was shown that these other groups do not work either in the two-gbit case. It seems reasonable to conjecture from our results and the results in~\cite{MMAPG2} that also for more gbits groups other than $\mathrm{SO}(d)$ fail to yield any non-trivial solution. Furthermore, $\mathrm{SO}(d)$ is the natural choice for generalizing the geometrical meaning of the Bloch ball for spin-$\frac 1 2$ particles to higher spatial dimensions, namely that the Bloch vector defines a direction in physical space.\\

Another route might be to drop tomographic locality, as in~\cite{DakicBrukner}. In fact, the $d=2$ Bloch ball corresponds to the quantum bit over the real numbers, and if we simply define the corresponding $n$-gbit state space to be the $2^n$-level quantum states over the reals, then this defines a model with interesting computational power (namely, equal to standard quantum computation), albeit one that does not satisfy tomographic locality. A similar construction can be performed for the $d=5$ case of quaternionic quantum theory~\cite{GraydonMaster} (but see the subtleties pointed out in~\cite{BGWshort,BGW}). The problem is, however, that these two cases are extremely special: building the composite state space uses the postulate that the result is supposed to be a Euclidean Jordan algebra. This assumption is not consistent with any of the other cases $d\not\in\{2,3,5\}$.

Furthermore, tomographic locality is a very natural postulate: it formalizes the idea that the whole is just composed of its parts and the relations between them. Other forms of state space composition would have to violate this intuition. Furthermore, they would have to violate the fact that states of composite systems can be described by tensors, a fundamental structural property of quantum theory with a myriad of physical consequences.

A possible way to drop tomographic locality despite these problems would be to instead assume (some version of) \emph{purification}~\cite{ChiribellaPurification,Chiribella,ChiribellaEntropy}. While purification has been very successful as a postulate of quantum theory, in particular, by illuminating how several characteristic properties of quantum theory can be understood directly via diagrammatic reasoning~\cite{ChiribellaPurification}, it is also very strong as a postulate. In fact, it is so strong that it is currently not clear whether there are any theories other than standard complex quantum theory and some of its subtheories~\cite{BarnumLee} that satisfy it. A potential alternative can be found in the work by Galley and Masanes~\cite{Galley1,Galley2}  who have pioneered an approach to construct composite state spaces directly in terms of group representations, without assuming tomographic locality.

We have made the implicit assumption that computations are carried out in the following way: first, the input is encoded into the initial state; then the actual computation is performed fully reversibly; and finally, the output is read out by a measurement. While this is arguably a natural standard scenario in the \emph{reversible} context, one might ask whether allowing measurements at any point during the computation could increase the computational capabilities of a theory. This is not the case in standard quantum mechanics, where all measurements can be modelled as unitary transformations on the system and an ancilla. But in principle, it might be true for other computational probabilistic theories.

Finally, one could drop the assumption of reversibility and/or connectedness of the groups, and consider transformations that are elements of some semigroup or finite group. However, dropping connectedness means giving up continuous (time) evolution, a large step away from our current conception of physics. Similarly, dropping reversibility means a substantial departure from our current understanding of fundamental physics: it would mean to give up conservation of information at the fundamental level.

\end{document}